\documentclass[aps,prb,twocolumn,superscriptaddress]{revtex4}
\usepackage{amsmath,amssymb}
\usepackage{graphicx}
\usepackage{epstopdf}
\usepackage{bm}
\usepackage{color}
\usepackage{mhchem}

\newcommand{\slr}{$T_1^{-1}$}

\newcommand{\ctob}{Cu$_2$Te$_2$O$_5$Br$_2$}

\newcommand{\ctox}{\ce{Cu$_2$Te$_2$O$_5$X$_2$}}

\begin{document}


\title{Persistence of singlet fluctuations in the coupled spin tetrahedra
system \ctob\ revealed by high-field magnetization and $^{79}$Br
NQR - $^{125}$Te NMR}


\author{S.-H. Baek}
\email[]{sbaek.fu@gmail.com}
\affiliation{IFW-Dresden, Institute for Solid State Research,
PF 270116, 01171 Dresden, Germany}
\author{K.-Y. Choi}
\affiliation{Department of Physics, Chung-Ang University, Seoul 156-756, Republic of Korea}
\author{H. Berger}
\affiliation{Institute de Physique de la Matiere Complexe, EPFL, CH-1015
Lausanne, Switzerland}
\author{B. B\"{u}chner}
\affiliation{IFW-Dresden, Institute for Solid State Research,
PF 270116, 01171 Dresden, Germany}
\affiliation{Institut f\"ur Festk\"orperphysik, Technische Universit\"at
Dresden, 01062 Dresden, Germany}
\author{H.-J. Grafe}
\affiliation{IFW-Dresden, Institute for Solid State Research,
PF 270116, 01171 Dresden, Germany}
\date{\today}

\begin{abstract}
We present high-field magnetization and $^{79}$Br nuclear quadrupole resonance
(NQR) and $^{125}$Te nuclear magnetic resonance (NMR)
studies in the weakly coupled Cu$^{2+}$ ($S=1/2$) tetrahedral system \ctob.
The field-induced level crossing effects were observed by
the magnetization measurements in a long-ranged magnetically ordered state which was
confirmed by a strong divergence of the spin-lattice
relaxation rate \slr\ at $T_0=13.5$ K. In the
paramagnetic state, \slr\ reveals
an effective singlet-triplet spin gap much larger than that observed by
static bulk measurements.
Our results imply that the inter- and the intra-tetrahedral
interactions compete, but at the same
time they cooperate strengthening effectively the local intratetrahedral
exchange couplings. We discuss that the unusual feature originates from the
frustrated intertetrahedral interactions.
\end{abstract}




\maketitle

Frustrated quantum spin systems have proven to be fertile ground for studying
the rich variety of quantum phases, novel magnetic excitations, and quantum
criticality.\cite{mila00}
In particular, in quasi-zero dimensional systems where localized spin clusters (dimers or
tetrahedra with a singlet ground state) are weakly
coupled to each other, combined quantum effects of reduced dimensionality and
frustration can lead to an intriguing ground state due to the proximity to a
quantum critical point (QCP). In this case, the concomitant
occurrence of localized singlet fluctuations and collective
gapless excitations may allow Goldstone-like and gapped transverse modes and a
longitudinal mode.\cite{lemmens03,gros03,ruegg08}

The oxohalide \ctox\ (X=Br,Cl) represents a weakly coupled spin
tetrahedral system in which four Cu$^{2+}$ spins form a distorted tetrahedron
with comparable nearest and next-nearest exchange constants, $J_1$ and $J_2$.\cite{johnsson00, 
valenti03} It undergoes an incommensurate magnetic ordering at 
11.4 K for Br and at 18.2 K for Cl,\cite{zaharko04} but the nature
of the magnetic behaviors for the two systems differs from each other.
The Cl system features an almost fully ordered magnetic moment of Cu$^{2+}$ 
(0.88 $\mu_B$)\cite{zaharko04,zaharko06} and a mean-field-like behavior of 
$T_N$ in fields.\cite{lemmens01} In this regard, the chloride can be approximated by a
classical magnet, although the effect of residual quantum fluctuations
seems to persist.\cite{choi09}
For the Br system, by comparison, an unusual transition at 11.4 K
(Ref. \onlinecite{lemmens01}) takes place in a singlet
background with a strongly reduced magnetic moment of $0.4 \mu_B$.\cite{zaharko06}
The anomalous field dependence of the transition 
temperature\cite{lemmens01,sologubenko04} $T_0$ and the
presence of a longitudinal magnon\cite{gros03,jensen03} coexisting with a gapped
singlet-like mode manifests the dominance of quantum fluctuations.
Chemical and hydrostatic pressure
measurements suggest the closeness of the Br system to a 
QCP.\cite{wang05,kreitlow05,crowe06,wang11b} 

Currently, there is a lack of understanding of the interplay between the inter- and
the intratetrahedral interactions in \ctob, which is likely the cause for
many discrepancies between experiment and theory in this system. Motivated by this, we
carried out high-field
magnetization measurements, as well as $^{79}$Br ($I=3/2$) nuclear
quadrupole resonance (NQR) and
$^{125}$Te ($I=1/2$) nuclear magnetic resonance (NMR) in a high-quality \ctob\
single crystal. A remarkable finding is that a softening of the
triplon spectral weight in this system is negligible, in
spite of the large intertetrahedral (IT) coupling which leads to a
three-dimensional (3D) magnetic order at low temperatures, indicating
the unconventional role of the IT coupling for the magnetic properties of
\ctob.

Single crystals of \ctob\ were prepared by the halogen vapor transport
technique, using TeBr$_4$ and Br$_2$ as transport agents.
High-field magnetization measurements were performed at the Dresden High Magnetic Field
Laboratory using a pulse magnet with a 20 ms-duration pulsed field. The
magnetic moment was detected by a standard inductive method with a pick-up
coil device up to 60 T at 1.4 K.
$^{79}$Br NQR and $^{125}$Te NMR measurements were
performed in the temperature range 13 -- 300 K.
A single NQR line of
$^{79}$Br was detected at 87.4 MHz at 15 K in zero
field, as reported in Ref.~\onlinecite{comment10}. $^{125}$Te NMR was
measured in the fields of 8 T and 11.8 T along the $c$ axis.
While the $^{125}$Te NMR spectrum consists of four
lines from four inequivalent Te sites for an arbitrary orientation of the
magnetic field,\cite{comment10} those
lines collapse into a very narrow single line when the external field $H$ is
parallel to the $c$
axis, allowing accurate determination of the Knight shift $\mathcal{K}$
and complete saturation of the line for the measurements of \slr.

\begin{figure}
\centering
\includegraphics[width=\linewidth]{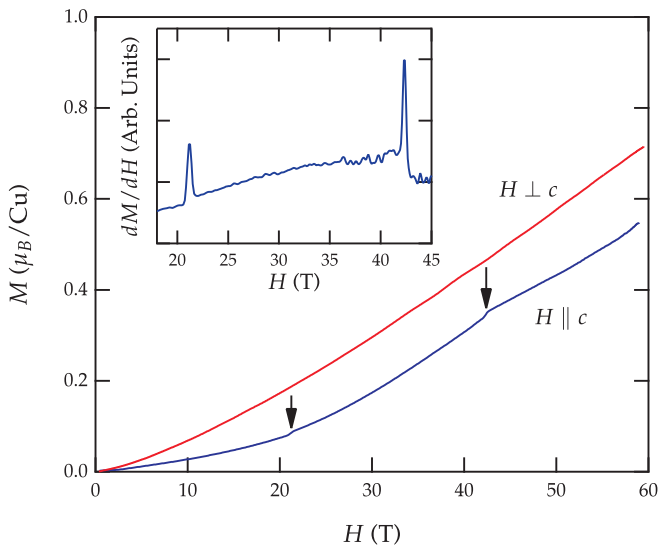}
\caption{\label{fig:mag} Magnetization $M$ versus pulsed magnetic field for
\ctob\ measured at 1.4 K in a field direction parallel and perpendicular
to the $c$ axis, respectively. Inset: derivative $dM/dH$ of the magnetization
for $H\parallel c$. Two kinks in the magnetization curve are evident by
sharp peaks in $dM/dH$. }
\end{figure}

Figure~1 shows the high-field magnetization curve $M(H)$ measured at 1.4 K in
the ordered state. A strongly anisotropic magnetization behavior was observed.
For $H$ perpendicular to the $c$ axis, $M(H)$ displays a linear
field dependence at high fields while it shows a smaller slope at low
fields. The change in the slope might be related to a
spin-flop-like transition for helical magnetic ordering. In contrast, for $H$
parallel to the $c$ axis, $M(H)$ is much reduced with a
concave curvature. Up to 60 T, we find two tiny magnetization jumps
at $H_{c1}=21.2$ T and $H_{c2}=42.4$ T, which are equally spaced as evident from the
derivative $dM/dH$.
We find that $g\mu_B H_{c1}=31$ K, where $g=2.15$,\cite{johnsson00} agrees
with the onset of the Raman
scattering continuum, i.e., 43 cm$^{-1}$ =  $2\Delta$ = 62 K, where $\Delta$
can be interpreted as the spin gap between the lowest excited triplet ($S_z=+1$)
and a singlet ($S=0$).\cite{lemmens01} Therefore, we conclude that the first level
crossing occurs at $H_{c1}$. Then, $H_{c2}$ may be attributed to the level
crossing between one of the
quintet ($S_z=+2$) and a singlet. From the slope of
$-2g\mu_B$, the quintet energy level at $H=0$ is estimated to be $\sim125$ K from the
singlet which also turns out to be a reasonable value (i.e., $\sim 3J_1$
where $J_1\sim 47$ K (Ref. \onlinecite{ruegg08}) is the nearest neighbor 
exchange) [see the red (gray) solid line in Fig. 4].
The finite slope in $M(H)$ below $H_{c1}$
indicates the presence of magnetic moments due to an admixture of triplets to
the otherwise singlet ground state. Dzyaloshinsky-Moriya (DM) interactions can 
admix triplet excitations 
into the singlet ground state and thus give rise to
a weak linear field dependence for $H < H_{c1}$. However, they are not sufficient
to explain the observed strong, nonlinear increase of the magnetization. Since
the ground state is magnetically ordered, we invoke the substantial IT
interactions as an origin. However, the observation of the magnetization steps is
compatible neither with a half-magnetization
plateau predicted for a chain of spin tetrahedra in a spin gap 
state,\cite{totsuka02} nor with a mean-field theory of coupled tetrahedra 
which does 
not show such steps.\cite{jensen03} This suggests an intriguing role of the IT
interactions, which induce a long-range ordering while retaining
the discrete energy levels of an isolated tetrahedron. This might be associated
with a complex frustrated IT interactions.

\begin{figure}
\centering
\includegraphics[width=\linewidth]{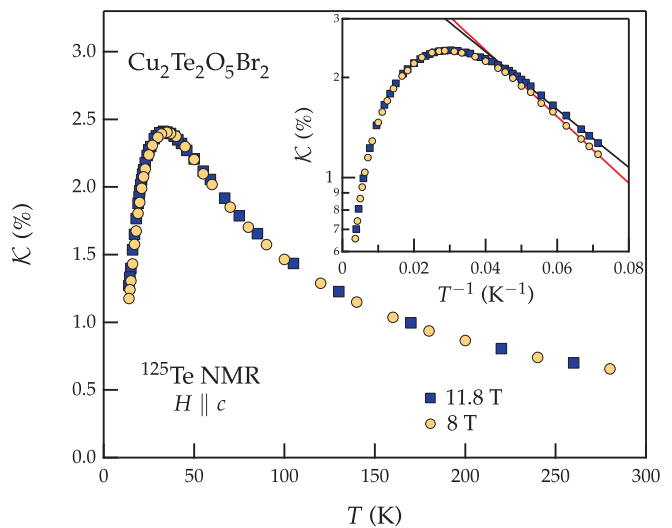}
\caption{\label{fig:K} Temperature dependence of the $^{125}$Te Knight shift
$\mathcal{K}$ measured in a single crystal of \ctob\ in external fields of 8
and 11.8 T along the $c$ axis.
$\mathcal{K}(T)$ exhibits a well-pronounced maximum at 34 K, below which a rapid
decrease is followed. Inset: the $^{125}$Te Knight shift at low temperatures
reveals the presence of a singlet-triplet gap which is suppressed with
increasing $H$. }
\end{figure}

Figure~2 shows the $^{125}$Te Knight shift $\mathcal{K}$ measured at 8 T and
11.8 T parallel to the $c$ axis as a
function of $T$. $\mathcal{K}(T)$ increases with decreasing $T$
reaching a maximum at $\sim 34$ K.  The local maximum is followed by a
rapid drop at low $T$,
which is a typical behavior found in spin gap systems.
In spin dimer and spin ladder systems,
$T^{\alpha}\exp(-\Delta_\mathcal{K}/T)$ is used to extract the singlet-triplet gap
of $\Delta_\mathcal{K}$.\cite{troyer94} However, no analytic expression for 
the exponent $\alpha$ is known 
for a spin tetrahedral system. In addition, in our case the prefactor $T^{\alpha}$
is expected to be nullified due to strong IT and DM interactions. Thus, an activated Arrhenius
form is employed to describe $\mathcal{K}(T)$ at low $T$.
Note that the bulk static susceptibility $\chi(T)$ was unable to measure the
spin gap directly since it is strongly
affected by paramagnetic impurities at low $T$,\cite{johnsson00} whereas 
$\mathcal{K}(T)$, i.e., the local static susceptibility is 
insensitive to impurities.
As shown in the inset of Fig.~2, $\mathcal{K}(T)$ depends on $H$ only
at sufficiently low $T$, yielding
$\Delta_\mathcal{K}=23$ K and 20 K, at 8 T and 11.8 T respectively.
The $H$-dependence of $\Delta_\mathcal{K}$ is in
satisfactory agreement with the Zeeman splitting
estimated from $H_{c1}$ in $M(H)$ (see the open triangles of Fig. 4).

\begin{figure}
\centering
\includegraphics[width=\linewidth]{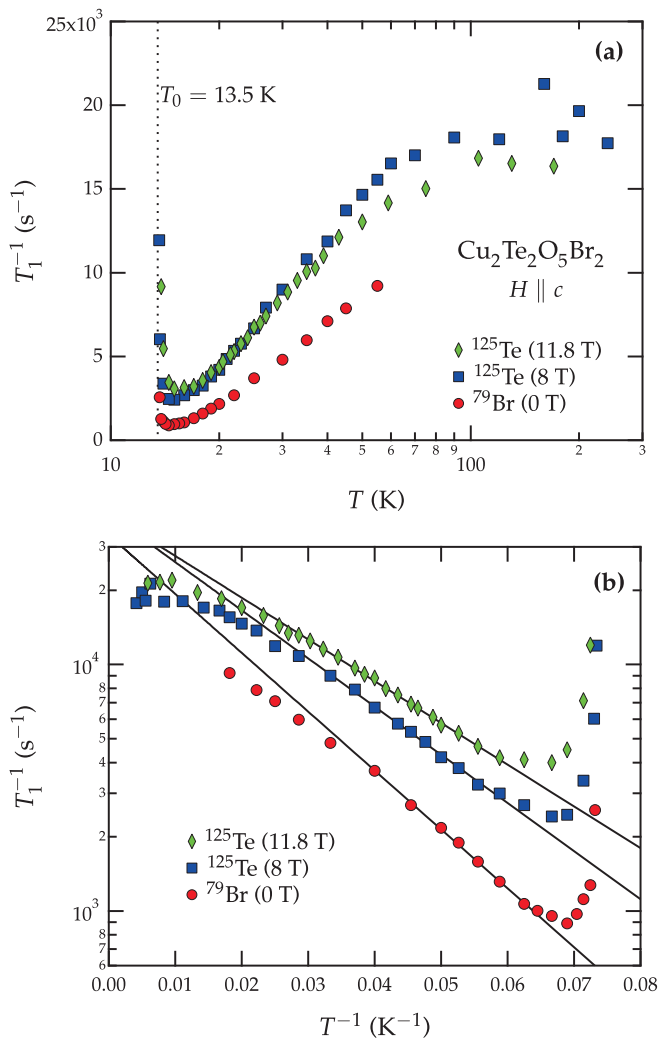}
\caption{\label{fig:T1} Temperature dependence of \slr\ measured via $^{79}$Br
NQR in zero field and $^{125}$Te NMR in
external fields of 8 T and 11.8 T. (a) \slr\ versus $T$.
\slr\ falls rapidly below $\sim 70$ K, showing the existence of a spin gap.
A strong divergence of \slr\ arises due to the 3D magnetic
order at $T_0=13.5$ K.
(b) \slr\ versus $T^{-1}$. Field-dependent activation behavior is clearly identified below
$\sim 30$ K.  Fits with an Arrhenius form (solid lines) yield
the $H$-dependent energy gap $\Delta_{T_1}(H)$. Data at 11.8 T were
offset vertically by $2\times 10^3$ s$^{-1}$ for clarity. }
\end{figure}

The spin-lattice relaxation rates \slr\ of $^{125}$Te in fields parallel to
the $c$ axis as well as
$^{79}$Br in zero field as a function of $T$ are presented in Fig. 3(a).
While \slr\ is almost $T$-independent above 100 K,
it starts to decrease exponentially with decreasing $T$ below $\sim 70$ K.
This indicates that most of spectral weights lie at
the spin singlet state due to proximity to a QCP, despite the long-ranged
magnetic ordering at $T_0$.
(For the $^{79}$Br, \slr\ is not measurable
above 50 K due to the shortening of the spin-spin relaxation time $T_2$.)
Near 15 K, all the \slr\ data, regardless of the presence or the magnitude of $H$, start to
increase abruptly and diverge at a well-defined temperature $T_0=13.5$ K,
confirming the magnetic origin of the transition.
The extremely narrow transition width ($\ll T_0$) above $T_0$ corroborates the 3D
character of the
magnetic order suggested in previous studies.\cite{jaglicic06,jensen09,comment10}
Surprisingly, $T_0$ identified in our study is considerably higher than 11.4 K in Ref.
\onlinecite{lemmens01} and 10.5 K in Ref. \onlinecite{comment10}.
While a higher $T_0$ usually suggests a higher
sample quality, the strongly sample-dependent variation of $T_0$ up to 30\% is
quite unusual. Rather, we interpret such a largely
varying $T_0$ as a signature that the system lies
in the vicinity of a QCP. Namely, $T_0$ is directly related to the quantum
instability which is very sensitive to nonmagnetic impurities or ``chemical
doping''.\cite{prester04}
Another peculiar feature is that $T_0$ is robust against $H$ up to 11.8 T.
This is in good agreement with the thermal conductivity measurements in which
$T_0$ increases with increasing $H$ only for $H \perp c$
but does not change for $H \parallel c$ up to 6 T.\cite{sologubenko04}
Thus, our data indicate that the anisotropy of $T_0(H)$ persists at least up to 12 T, which
may be related to the anisotropy of the magnetization.

\begin{figure}
\centering
\includegraphics[width=\linewidth]{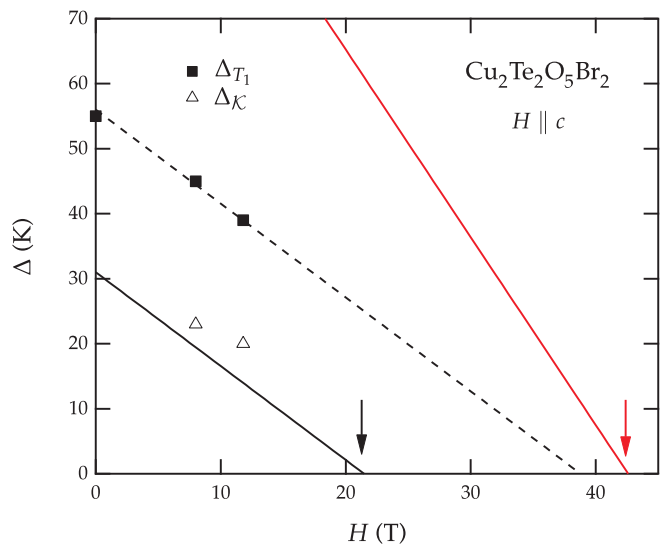}
\caption{\label{fig:delta} Spin gaps $\Delta_{T_1}$ and $\Delta_\mathcal{K}$ as a function
of $H$ determined from
\slr\ and $\mathcal{K}$, respectively. Solid and dashed lines which are
parallel each other are the Zeeman
splitting for the singlet-triplet gap assuming a $g$-factor of 2.15.
Arrows denote the critical fields in the magnetization.
The red (gray) solid line is an estimate due to the quintet state separated by
$\sim125$ K from the singlet ground state without field. }
\end{figure}

In Fig. 3(b), \slr\ is plotted against $T^{-1}$ to examine
a thermal activation behavior. Below $\sim 30$ K (0.033 K$^{-1}$), all the
data are well fit by an Arrhenius form
$T_1^{-1}\propto \exp(-\Delta_{T_1}/T)$ giving rise to the $H$-dependent
spin gap $\Delta_{T_1}(H)$ (solid lines).
The obtained values of the gap $\Delta_{T_1}$ as a function of $H$ are
drawn in Fig. 4.
Clearly, $\Delta_{T_1}(H)$ follows a Zeeman splitting expected for a
singlet-triplet gap in an isolated tetrahedron, shown as a dashed
line given by $\Delta_{T_1}(0)-g\mu_B H$ with $\Delta_{T_1}(0)=56$ K.
Thus, our results show that $H$ parallel to the
$c$ axis has no influence on both $T_0$
and the spin gap itself, and does not cause any field-induced effect, suggesting that the
singlet tetrahedron is almost intact in fields.  This is in contrast to the theoretical
prediction\cite{kotov04} and thus suggests that the
DM interaction, which should give a nontrivial $H$-dependence,
is fairly small in this system, at least, for $H\parallel c$.

An unexpected finding is that the spin gap values of $\Delta_{T_1}(H)$ are much
bigger than those from both the magnetization and the Knight shift, implying that
\slr\ sees a larger spin gap. In fact, a
larger gap from \slr\ than from the spin susceptibility is
often observed in low-dimensional spin gap systems.\cite{shimizu95,furukawa96} 
One plausible explanation is because \slr\ 
samples the $q$-sum dynamical susceptibility, i.e.,
$T_1^{-1}\propto T\sum_q A^2(q)\chi''(q,\omega_0)$ where $A(q)$ is the hyperfine
form factor and $\omega_0$ the nuclear Larmor frequency, so that
$\sum_q\chi''(q,\omega_0)$ may exhibit a larger gap if the gap
formed in $\chi''(q,\omega_0)$ is larger at $q=Q$ than at $q=0$.
That is, the contribution of the dominant spin
fluctuations at $q=Q$ to \slr\ decreases more rapidly than
that near $q=0$ with decreasing $T$, resulting in a larger gap.
However, such a strong $q$-dependence of the spin gap is somewhat unlikely in the present
case because the singlet-triplet gap should correspond to band-like gapped 
excitations as detected in inelastic neutron scattering (INS) 
study.\cite{crowe05}  
Instead, we note that $\Delta_{T_1} (H=0)=56$ K = 4.82 meV
falls into the center of the
gapped excitations in INS data.  Therefore, it seems that both \slr\ and INS, which are
commonly described by $\chi''(q,\omega)$,
detect similarly a spin gap larger than those obtained from static bulk measurements.
Also, since $\Delta_{T_1}$ clearly displays a Zeeman splitting for a
singlet-triplet gap, one can rule out a possible contribution to
$\Delta_{T_1}$ from the quintet state.

Then, we conjecture that the discrepancy in the measured spin gap values is due
to frustrated IT quantum fluctuations and anisotropic DM interactions. This 
may lead to a band-like broadening of triplet excitations.  Indeed, the recent 
single crystal INS measurements show that most of the spectral weight of 
excitations remains gapped without a substantial softening to a lower energy.\cite{prsa09}
In this case, the spin gap determined by the Raman scattering, the
magnetization, and the Knight shift
corresponds to the lowest energy of the triplon band.
On the other hand, \slr, i.e., the $q$-sum dynamical susceptibility, may
see the \textit{effective} spin gap where most of the spectral weight remains 
gapped.  
Assuming that the singlet-triplet gap in a
completely decoupled tetrahedron $\Delta_\text{ST} = J_1 = 47$ K,\cite{gros03}
$\Delta_{T_1}(0)=56$ K indicates that
$\Delta_\text{ST}$ is promoted,
rather than suppressed, by $J_\text{IT}$.
The shift of the spectral weight to higher than $\Delta_\text{ST}$ highlights 
an unconventional role of the IT interactions.  In semiclassical theories, 
higher dimensional interactions lead to a softening of the spectral weight.  
Thus, the gap difference $\Delta_{T_1}(0)- g\mu_B H_{c1} =  25 \text{ K }$
is related to the magnitude of $J_\text{IT}$, which in turn determines the 
triplon bandwidth and the longitudinal magnon energy.

The large
effective singlet-triplet spin gap detected by \slr\ in the paramagnetic state,
and the quantized energy levels in the
magnetically ordered state detected by the equally spaced magnetization jumps,
imply that the average spectral
weight of the triplon is shifted to higher energies while
the $q=0$ spectral weight remains intact through the magnetic transition.
This is contrasted by a coupled spin dimer system\cite{ruegg08}
and suggests the significance of the IT quantum fluctuations related to a zero 
dimensionality.
As a possible origin we resort to peculiar
spin networks. In the studied compound,
the IT interactions are frustrated since they
couple the four tetrahedra in the vertical, horizontal, and diagonal 
directions.\cite{valenti03}
In this unique spin network, the IT interactions can induce
the helical magnetic ordering without accompanying a softening of the triplon
spectral weight. Instead, an overall energy scale of the triplon
can be shifted to higher energies because the frustrated IT interactions
are added to the intratetrahedral ones. Here, frustration together with a zero 
dimensionality retains singlet fluctuations.  
Indeed, this accounts for the unconventional increase of the spin gap by
applying pressure, leading to a quantum
phase transition to a spin singlet state.\cite{kreitlow05,wang11b}
That is, both the magnitude and the frustration degree of the IT
interactions are enhanced by pressure.

In conclusion, a combined study of the magnetization and $^{125}$Te NMR-$^{79}$Br
NQR in the weakly coupled quantum spin system \ctob\ showed that
a 3D magnetic order emerges from
a singlet background which would be expected in a simple isolated spin $1/2$
tetrahedral system.
Remarkably, our data suggest that the IT
coupling not simply induces a 3D magnetic order but also
increases the effective spin gap by enhancing the intratetrahedral spin-exchange processes.
This unusual feature is attributed to the frustrated IT interactions which may account
for the discrepancy between experiment and theory,\cite{prsa09,jensen09}
and thus our findings will forward the establishment of a theoretical model
adequate for this unique quantum system.

We acknowledge S.-L. Drechsler and V. Kataev for useful discussions, and
Y. Skourski for a technical assistance in magnetization measurements.
KYC thanks EuroMagNET II under the EC Contract 228043 and the Korean NRF
(2009-0093817).

\bibliography{mybib}

\end{document}